\documentclass[referee]{aa} 

\usepackage{epsf}
\usepackage[dvips]{epsfig}
\usepackage{txfonts}
\usepackage{natbib}

\begin{document}

\title{ An Exact Equilibrium Model of an Unbound Stellar System in a
  Tidal Field} 
\titlerunning{Unbound Stellar System}

\author{Michael Fellhauer\inst{1} \and Douglas C. Heggie\inst{2}}
\authorrunning{M. Fellhauer \& D.C. Heggie}

\institute{Sternwarte Universit\"at Bonn, Germany \and School of
  Mathematics, University of Edinburgh, Scotland, UK} 

\date{Received / Accepted }

\abstract{
  Star clusters and dwarf galaxies gradually dissolve as they move in
  the potential of their host galaxy.  Once their density falls below
  a certain critical density (which is comparable with the background
  density of the galaxy) it is often assumed that their evolution is
  completed.  In fact the remnant of such a system forms a
  distribution of stars which are unbound to each other and which move
  on similar orbits in their host potential.
  With this motivation we study the evolution of an idealised unbound
  system and follow its expansion and dissolution in the tidal field
  of a model galaxy.  Initially the stars are uniformly distributed
  (with a density below the critical density) within an ellipsoidal
  volume.  The system itself travels on a circular orbit within a
  galaxy modelled as an isothermal sphere.  The initial velocities of
  the stars are chosen by assuming that they move on
  (three-dimensional) epicycles with guiding centre at the centre of
  the ellipsoid, though the usual epicyclic theory is altered to
  account for the self-gravity of the system.  This is believed to be
  the first exact equilibrium model of a stellar system in a tidal
  field. 
  Our main task is to study the stability of such configurations and
  the time-scale of their dissolution, as a function of the initial
  density and size of the ellipsoid.  If the time of dissolution is
  measured by an increase of the half-mass radius of $50$\%, we find
  that systems of low density ($\sim 1$\% of the background density)
  and small radius ($50$~pc on an orbit of radius $10$~kpc) can
  survive for about $20$ galactic orbits.  For small systems we show
  that the lifetime is approximately proportional to the inverse
  square root of the density.
}

\maketitle

\section{Introduction}
\label{sec:intro}

In recent years, considerable effort has been spent in examining the
destructive processes which lead to the dissolution of star clusters
and dwarf galaxies, and the influence of these processes on the
distribution function of these objects \citep[e.g.\ ][]{bau98} and on 
the host galaxy.  There are two main areas of interest.  First there
is the dissolution caused by internal processes such as evaporation
\citep[e.g.\ ][]{gie97,bau03}.  The second issue is the merging of
these objects with the body of the galaxy \citep[e.g.\ ][]{vel99} or
with  each other \citep{fel02}.  Within the context of the first
problem, investigations usually stop when the object has dissolved
according to some tidal stability criterion (Sec.~\ref{sec:setup}).
So far little attention has been paid to their subsequent fate,
i.e. what happens to these objects after they become unbound entities. 

Retrograde motions are expected to play an important role in these
unbound systems.  It has been known for a long time \citep{hen69}
that stars in retrograde motion about a star cluster (at least for
clusters on circular galactic orbits) are stable (i.e.\ orbitally
stable) even outside the tidal boundary; their orbits are best thought
of as perturbed epicycles.  As a star cluster loses mass and its tidal
radius shrinks, we may expect that it is surrounded by some stars
moving approximately on such orbits.

Based on these considerations, the present study starts with a very
special toy model in which all the stars move on retrograde orbits
resembling epicycles with a common guiding centre.  Their spatial
distribution is a homogeneous ellipsoid, and the centre of the
ellipsoid moves on a circular orbit within the potential of an
isothermal sphere, which is taken as a model of the host galaxy.  The
simplicity of our setup permits some analytical investigation which
gives useful insights into the results of numerical simulations.  It
is also believed to be the first example of an exact equilibrium of a
stellar system in a tidal field. 

Having set up the sequence of models, our main interest turns to their
stability.  In particular, we aim to investigate the time-scales on
which they evolve and finally disperse, and their spatial and
kinematic properties as they do so.

In the next section we briefly go through the theory of epicyclic
motion with modifications to account for the self-gravity of the
system itself.  We also explain the setup of our toy model.  Then we
describe the results of our simulations, which cover a large parameter
space.  Finally we give our conclusions, and discuss their possible
relevance for real systems, such as those found in and around our
Milky Way.

\section{Theory \& Setup}
\label{sec:setup}

{\bf
The basic idea of our model is the following.  Inside an ellipsoid of
uniform density in a linear tidal field, orbits are modified
epicycles, with a harmonic motion in $z$ of amplitude $z_0$, say (c.f
Fig.~\ref{fig:orbit}).  We consider epicycles which are centred at the
centre of the ellipsoid.  For a given epicyclic amplitude, we
construct a distribution of $z_0$ so that the space density of this
superposition of epicycles is independent of $z$ (up to the edge of
the ellipsoid).  Then we construct a distribution of epicyclic
amplitudes so that the space density is also uniform in $x$ and $y$.
Because we have arranged for the amplitude of the epicyclic $x-$ and
$y-$ motions to be in the same proportion as the corresponding axes of
the ellipsoid, we can build up the entire ellipsoid in this way.

In principle this gives rise to an equilibrium model.  Our $N$-body
simulations, however, do not maintain equilibrium because of
relaxation, because the orbits are dynamically unstable, and because
the $N$-body simulations use the exact tidal field and not a linear
approximation. 

\begin{figure}[h!]
  \centering
  \epsfxsize=8cm
  \epsfysize=8cm
  \epsffile{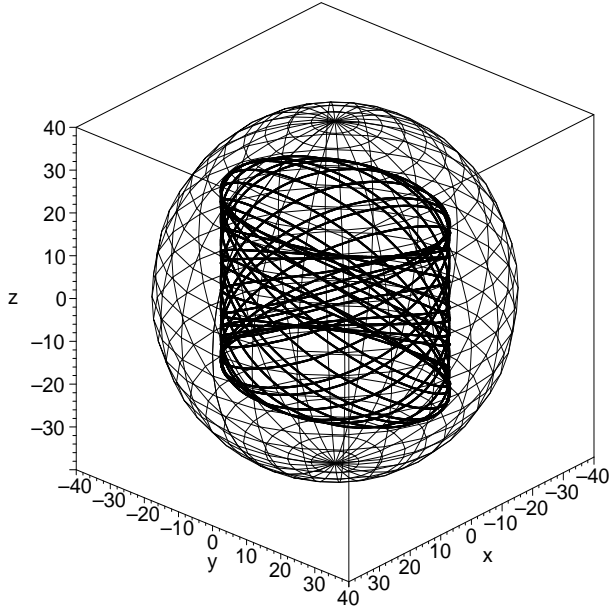}
  \caption{A schematic representation of a uniform ellipsoid and one
    orbit.} 
  \label{fig:orbit}
\end{figure}
}

Let $X,Y,Z$ be galactocentric coordinates, with $Z$ perpendicular to
the plane of motion of the system under study, and let $R^2 = X^2 +
Y^2$.  Epicyclic theory is a linearised approximation in which stars
move in a potential 
\begin{eqnarray}
  \label{eq:theo}
  \Phi_{} & = & \frac{1}{2} \left. \frac{\partial^{2}
        \Phi_{\rm eff}} {\partial R^{2}} \right|_{D,0}  (R-D)^{2}
        + \frac{1}{2} \left. \frac{\partial^{2} \Phi_{\rm eff}}
        {\partial Z^{2}} \right|_{D,0} Z^{2}
\end{eqnarray}
\citep{bin87}, where $D$ denotes the distance to the axis of symmetry
of the galaxy and $\Phi_{\rm eff}(R,Z)$ is the effective galactic
potential.  We set 
\begin{eqnarray}
  \label{eq:theo2}
  \left. \frac{\partial^{2} \Phi_{\rm eff}} {\partial R^{2}}
  \right|_{D,0} & = & \kappa^{2}, \\
  \label{eq:theo3}
  \left. \frac{\partial^{2} \Phi_{\rm eff}} {\partial Z^{2}}
  \right|_{D,0} & = & \nu^{2},
\end{eqnarray}
where $\kappa$ is the epicyclic frequency.  For the potential of a
singular isothermal sphere these frequencies are given by
\begin{eqnarray}
  \label{eq:epifreqold1}
  \kappa & = & \sqrt{2}  \Omega, \\
  \label{eq:epifreqold2}
  \nu    & = & \Omega,
\end{eqnarray}
where $\Omega=V_{0}/D$ is the angular velocity of motion on a circular
orbit of radius $D$ around the host galaxy and $V_{0}$ is the circular
velocity. 

We adopt initial conditions in which stars are distributed uniformly
within an ellipsoid with semi-major axes $a, \xi a, a$, where $a>0$ is
a free parameter, and $\xi>1$ will be determined from the shape of
closed epicycles.  Therefore we have to take the potential of a
prolate ellipsoid of uniform density $\rho$ into account.  We use
rotating, cluster-centered coordinates $x$, $y$, $z$ with origin at
the centre of the ellipsoid of stars.  The coordinate axes are
directed away from the galactic centre, in a direction opposite that
of the circular motion round the galaxy, and orthogonal to the plane
of motion, respectively.  The appropriate differential equations in
the epicyclic approximation are then \citep{cha42}
\begin{eqnarray}
  \label{eq:epidiff1}
  \ddot x + 2 \Omega \dot y + (\kappa^2 - 4 \Omega^2) x  & = & 
  - B_{1} x  \\
  \label{eq:epidiff2}
  \ddot y - 2 \Omega \dot x \phantom{\ + (\kappa^2 - 4\Omega^2)x}
  & = & - B_{3} y \\
  \label{eq:epidiff3}
  \ddot z \phantom{+ 2 \Omega\dot y + (\kappa^2 - 4} + \nu^2 z &
  = & - B_{1} z .
\end{eqnarray}
Here, $B_{1}= 2\pi G \rho A_{1}$ and $B_{3}= 2\pi G \rho A_{3}$, where
$A_{1}$ and $A_{3}$ are given by standard formulae for the potential
of a homogeneous prolate ellipsoid \citep[][ Table 2-1]{bin87}.

This additional acceleration alters the frequencies of motion and the
shape of closed epicycles.  Eigensolutions of these equations give
frequencies $\kappa^{\prime}$ and $\nu^{\prime}$ defined by
\begin{eqnarray}
  \label{eq:eqiparanew1}
  \kappa^{\prime 2} & = & \frac{1}{2} \left( (\kappa^{2} + B_{1} 
      + B_{3}) \pm \sqrt{ (\kappa^{2} + B_{1}-B_{3})^{2}
      + 16\Omega^{2}B_{3} } \right) \\
  \label{eq:eqiparanew2}
  \nu^{\prime 2} & = & \nu^{2} + B_{1} 
\end{eqnarray}
and the epicyclic ratio 
\begin{equation}
  \label{eq:eqiparanew3}
  \xi  =  \frac{\kappa^{\prime 2} +
    \kappa^{2} - B_{1}} {2 \Omega \kappa^{\prime}} \ = \ \frac{2
      \Omega \kappa^{\prime}} {\kappa^{\prime 2} - B_{3}}
\end{equation}
Note incidentally that the two expressions for $\xi$ in
eq.(\ref{eq:eqiparanew3}) are identical if eq.(\ref{eq:epifreqold1})
and the upper sign in eq.(\ref{eq:eqiparanew1}) are used.

If $\vert \kappa \vert \gg B_1, B_3$, then eq.(\ref{eq:eqiparanew1})
implies that $\kappa^\prime\simeq\kappa$ (if we choose the upper
sign), and eq.(\ref{eq:eqiparanew3}) gives approximately the familiar
axial ratio of epicyclic motion.  In the general case, since we also
use $\xi$ for the axial ratio of the ellipsoid, the right side of
eq.(\ref{eq:eqiparanew3}) depends on $\xi$.  Therefore we have solved
for $\kappa^{\prime}$ and $\xi$ using the above equations iteratively
(taking the upper sign in eq.(\ref{eq:eqiparanew1})).

A second solution of the differential equations
(eqs.(\ref{eq:epidiff1}) and (\ref{eq:epidiff2})) is obtained if one
takes the lower sign for $\kappa^{\prime}$.  It leads to an imaginary
frequency $i k = \kappa^{\prime\prime}$ provided that
\begin{eqnarray}
  \label{eq:imagcrit}
  \rho & < & \frac{4 \rho_{0}}{A_{1}},
\end{eqnarray}
where $\rho_{0}$ denotes the density corresponding to the isothermal
background.  (This is always fulfilled in our parameter range, though
there is a series of {\sl bound} systems (i.e. systems not satisfying
eq.(\ref{eq:imagcrit})) where all frequencies are real.)  We can see
that $k\ll\kappa^{\prime}$ if $\rho$ is much smaller than the critical
density (cf.  eq.(\ref{eq:imagcrit})), and then we have: 
\begin{eqnarray}
  \label{eq:kappaimg}
  k^{ 2} & = & -\frac{1}{2} \left( (\kappa^{2} +
  B_{1} + B_{3}) - \sqrt{ (\kappa^{2} + B_{1}-B_{3})^{2}
    + 8\kappa^{2}B_{3} } \right) \nonumber \\
  & \approx &  B_{3} \\
  \label{eq:xiimg}
  \xi^{\prime} & = & \frac{2 \Omega k}{ {k^2} + B_{3}} \nonumber \\
  & \approx & \frac{\Omega} {\sqrt{B_{3}}}, 
\end{eqnarray}
in which $\xi^\prime$ is the ratio of the components of the
corresponding eigenvector. 

Within the epicyclic approximation the complete solution is a linear
combination of both solutions, i.e. 
\begin{eqnarray}
  \label{eq:epimot1}
  x(t) & = & x_{0} \cos(\kappa^{\prime} t + \psi)  + \lambda
  \exp(kt) \\
  \label{eq:epimot2}
  y(t) & = & \xi x_{0} \sin(\kappa^{\prime} t + \psi)  +
  \lambda \xi^{\prime} \exp(kt) \\
  \label{eq:epimot3}
  z(t) & = & z_{0} \cos(\nu^{\prime} t + \phi) 
\end{eqnarray}
where $x_{0}$, $z_{0}$, $\psi$, $\phi$ and $\lambda$ are free
parameters. (As with any unstable equilibrium when there is no
dissipation, there is also an exponentially decaying solution, which
we have ignored.  It corresponds to the stable invariant manifold of
the epicyclic motion, and is of no practical importance, being of
measure zero.) 

As already mentioned the starting configuration is a homogeneous
distribution of particles in an ellipsoid with axes $a$, $\xi a$ and
$a$.  To achieve the correct spatial distribution we choose the
parameters $x_{0}$ and $z_{0}$ according to the following
distributions  
\begin{eqnarray}
  \label{eq:distfunc1}
  F(x_{0}) & = & 1 - \left( 1 - \frac{x_{0}^{2}}{a^{2}}
  \right)^{3/2}, \\ 
  \label{eq:distfunc2}
  F(z_{0}) & = & 1 - \sqrt{ 1 - \frac{z_{0}^{2}}
  {a^{2}-x_{0}^{2}}},
\end{eqnarray} 
with $x_{0} \in [0,a]$ and $z_{0} \in [0, \sqrt{a^{2} - x_{0}^{2}}]$.
The parameters $\psi$ and $\phi$ are chosen randomly in the interval
$[0,2\pi]$. 

{\bf
Our recipe for constructing this model is given by
eqs.(\ref{eq:epimot1})-(\ref{eq:epimot3}) (with three similar
equations for the initial velocities and adopting $\lambda = 0$ for
the initial setup) and eqs.(\ref{eq:distfunc1}) and
(\ref{eq:distfunc2}).  Here we show that this gives rise to the
desired uniform space density.  

From eqs.(\ref{eq:distfunc1}) and (\ref{eq:distfunc2}) it follows that
the joint probability density of the positive quantities $x_{0},
z_{0}$ is  
\begin{eqnarray}
  \label{eq:distfunc3}
  f(x_{0},z_{0}) & = & \frac{3 x_{0} z_{0}}{a^{3}} (a^{2} - x_{0}^{2}
  - z_{0}^{2})^{-1/2},
\end{eqnarray}
over the domain $x_{0} > 0, z_{0} > 0, x_{0}^{2} + z_{0}^{2} < a^{2}$.
We also take the distributions of $\psi, \phi$ to be independent and
uniform on $[0,2\pi]$.  The probability density of $x,y,z$ is
therefore 
\begin{eqnarray}
  \label{eq:distfunc4}
  f(x,y,z) & = & \frac{3} {4 \pi^{2} a^{3}} \int x_{0} z_{0}
  (a^{2} - x_{0}^{2} - z_{0}^{2})^{-1/2} \times \nonumber \\ 
  & & \delta (x - x_{0} \cos \psi) \delta (y - \xi x_{0} \sin \psi) 
  \delta (z - z_{0} \cos \phi) dx_0 dz_0 d\psi d\phi. 
\end{eqnarray}
The integration can be readily carried out by thinking of $x_0,\psi$
as a pair of plane polar coordinates and transforming to the
corresponding cartesian coordinates; similarly with $z_0,\phi$.  The
result is $f(x,y,z) = 3 / (4\pi\xi a^3)$, which is just the inverse of
the volume of an ellipsoid with semi-axes $a, \xi a, a$, as required.

Since the evolution of position with time is just a translation in
$\psi$ and $\phi$, it is clear that the space density remains uniform
on this ellipsoid.
}

The above coordinates and velocities are defined in the
cluster-centred frame which moves around the galaxy.  To transform to
the galactocentric rest-frame, we use the equations 
\begin{eqnarray}
  \label{eq:coordtrafo1}
  X(t) & = & (D+x) \cos(\Omega t) + y \sin(\Omega t) \\ 
  \label{eq:coordtrafo2}
  Y(t) & = & - (D+x) \sin(\Omega t) + y \cos(\Omega t)  \\ 
  \label{eq:coordtrafo3}
  Z(t) & = & z \\
  \label{eq:coordtrafo4}
  \dot{X}(t) & = & (\dot{x}+y\Omega) \cos(\Omega t) +
  (\dot{y}-(D+x)\Omega) \sin(\Omega t) \\
  \label{eq:coordtrafo5}
  \dot{Y}(t) & = & - (\dot{x}+y\Omega) \sin(\Omega t) +
  (\dot{y}-(D+x)\Omega) \cos(\Omega t) \\
  \label{eq:coordtrafo6}
  \dot{Z}(t) & = & \dot{z} 
\end{eqnarray}

To ensure that the initial distribution is not bound in the usual
sense, its density must lie below a certain critical density.  Roughly
speaking this is the density $\rho_{0}$ of the background, i.e. the
density which causes the galactic potential, and so
\begin{eqnarray}
  \label{eq:rho}
  \rho & < & \frac{V_{0}^{2}}{4 \pi G D^{2}} \ = \ \rho_{0}
\end{eqnarray}
(cf. eq.(\ref{eq:imagcrit})).

The simulations were performed with the particle-mesh code {\sc
  Superbox} \citep{fel00} with 1,000,000 particles and meshes with
$n=128$ grid points per dimension, to achieve high resolution.  We
chose the following grid sizes for the three nested grids which the
code uses: $R_{\rm system}=12$~kpc, $R_{\rm   out}=2$~kpc and $R_{\rm
  core}=0.4$~kpc.  As the resolution of {\sc   Superbox} is given by
the length of one grid cell, these parameters give a resolution of
$6.45$~pc (${\rm res} = 2R_{\rm core}/(n-4)$) in the central region of
the particle distribution.  The code has a simple leapfrog integration
scheme with fixed timestep, which in our simulations was DT$=0.1$~Myr.
To ensure that our conclusions do not depend on the parameters of the
code, we repeated one simulation with lower resolution and lower
particle number, and found no significant change in the results.

\section{Results}
\label{sec:results}

Using initial conditions specified in the previous section, we have
conducted a parameter survey for various choices of the semi-minor
axis $a$ and the density $\rho$ (Tab.~\ref{tab:parameter}).  The
distance to the galactic centre and the circular velocity were held at
$D=10$~kpc and $220$~kms$^{-1}$, respectively.

\begin{table}
  \begin{center} 
    \caption{Parameter space of the simulations.  Each simulation
      performed is marked with a bullet point.}  
    \label{tab:parameter}
    \begin{tabular}[h!]{r|rrrrrrrr} \hline 
      $a$ & \multicolumn{8}{c}{$\rho/\rho_{0}$} \\ 
      $[$pc$]$ & 0.01 &
      0.05 & 0.10 & 0.50 & 1.00 & 2.00 & 5.00 & 10.00 \\
      \hline \hline 10 & & & $\bullet$ & & & & & \\ 50 & $\bullet$ &
      $\bullet$ & 
      $\bullet$ & $\bullet$ & $\bullet$ & $\bullet$ & $\bullet$ &
      $\bullet$ \\ 
      100 & & & $\bullet$ & & &  & & \\ 
      500 & $\bullet$ & $\bullet$ & $\bullet$ & $\bullet$ & $\bullet$ &
      & $\bullet$ & $\bullet$ \\ 
      1000 & $\bullet$ & $\bullet$ & $\bullet$ & $\bullet$ & $\bullet$ &
      & $\bullet$ & $\bullet$ \\ 
      5000 & & & $\bullet$ & & & & & \\ \hline 
    \end{tabular}
  \end{center}
\end{table}

\subsection{Lifetime of the systems}

\begin{figure}
  \begin{center} 
    \epsfxsize=8cm 
    \epsfysize=8cm
    \epsffile{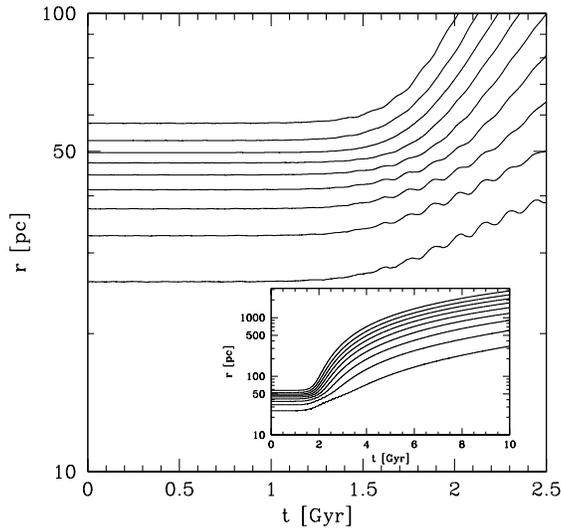} 
    \caption{Lagrangian radii (10, 20, ..., 90\%) for the simulation
      with $a=50$~pc and $\rho/\rho_{0}=0.1$.  The large figure shows
      the time interval during which the distribution is stable; the
      small figure shows the whole simulation.} 
    \label{fig:lagrad} 
  \end{center}
\end{figure}

In every simulation which satisfied eq.~(\ref{eq:rho}) initially, the
distribution was quite stable for several orbits around the galactic
centre.  Fig.~\ref{fig:lagrad} shows the Lagrangian radii of the
simulation where $a=50$~pc and $\rho/\rho_{0}=0.1$.  The orbital
period around the galaxy in our simulations was about $280$~Myr.
Clearly the distribution did not evolve much for almost $6$
revolutions around the galactic centre.

\begin{figure}
  \begin{center} 
    \epsfxsize=8cm 
    \epsfysize=8cm
    \epsffile{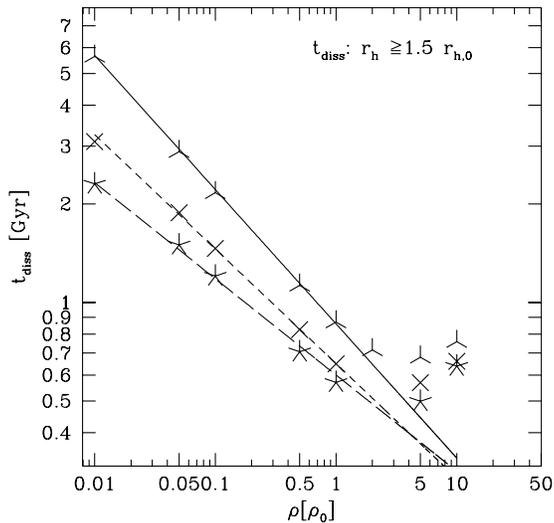} 
    \caption{Dissolution time as a function of the initial density of
      the particle distribution for different values of $a$.
      Three-pointed stars correspond to $a=50$~pc, crosses to
      $a=500$~pc and five-pointed stars to $a=1000$~pc.  The
      dissolution time is a decreasing function of $\rho$, except for
      $\rho/\rho_{0} > 1.0$.  In this case the distribution is indeed
      unstable (according to our criterion), but before expanding it 
      undergoes a slight initial collapse which leaves behind a small
      bound core.}  
    \label{fig:diss1} 
  \end{center} 
\end{figure}

Because this is an unbound configuration there is no conventional
definition for its time of dissolution.  For the purpose of comparing
the simulations, we adopt the time, $t_{\rm diss}$,  when
\begin{eqnarray}
  \label{eq:disscrit}
  r_{h} & = & 1.5 r_{h,0}
\end{eqnarray}
where $r_{h}, r_{h,0}$ denote the current and initial half-mass radii,
respectively.  With this definition, the dissolution time is shown in
Fig.~\ref{fig:diss1} as a function of the density, for different
values of $a$.  When $\rho$ is negligible, the particles describe
independent epicycles with a common guiding centre, and so $t_{\rm
  diss}$ is essentially infinite (in the epicyclic approximation). 

Except at the highest densities the results can be fitted
approximately with power laws 
\begin{eqnarray}
  \label{eq:powerfit}
  t_{\rm diss} = A \left(\frac{\rho} {\rho_{0}} \right)^{-\gamma}.
\end{eqnarray}
Best-fitting values for the parameters are shown in
Tab.~\ref{tab:powerfit}. 

\begin{table}
  \begin{center} 
    \caption{Fitting parameters for the power law relating the
      dissolution time to the initial density of the system.}
    \label{tab:powerfit} 
    \begin{tabular}[h!]{rcc} \hline 
      $a$ [pc] & $A$ [Gyr] &
      $\gamma$ \\ \hline \hline 50 & 0.86 & 0.41 \\ 500 & 0.65
      & 0.35 \\ 1000 & 0.60 & 0.29 \\ \hline 
    \end{tabular}
  \end{center}
\end{table}

We have shown in eq.~(\ref{eq:kappaimg}) that the exponent of the
exponential growing solution, $k$, is approximately equal to the
square root of $B_{3}$ which itself is proportional to the square-root
of the density.  This approximation is valid for very small initial
density.  If, however, we solve eq.(\ref{eq:kappaimg}) numerically
over the range of densities corresponding to our simulations, and fit
the results by a power law, we obtain approximately 
\begin{eqnarray}
  \label{eq:kapprop}
  k & \propto & \rho^{0.40 \pm 0.03}.
\end{eqnarray}
This shows that, as long as the size $a$ of the system is sufficiently
small that the epicyclic approximation holds, the dissolution time
should be almost proportional to $\rho^{-0.4}$.  This is confirmed by
Tab.~\ref{tab:powerfit}.

\begin{figure}
  \begin{center} 
    \epsfxsize=8cm 
    \epsfysize=8cm
    \epsffile{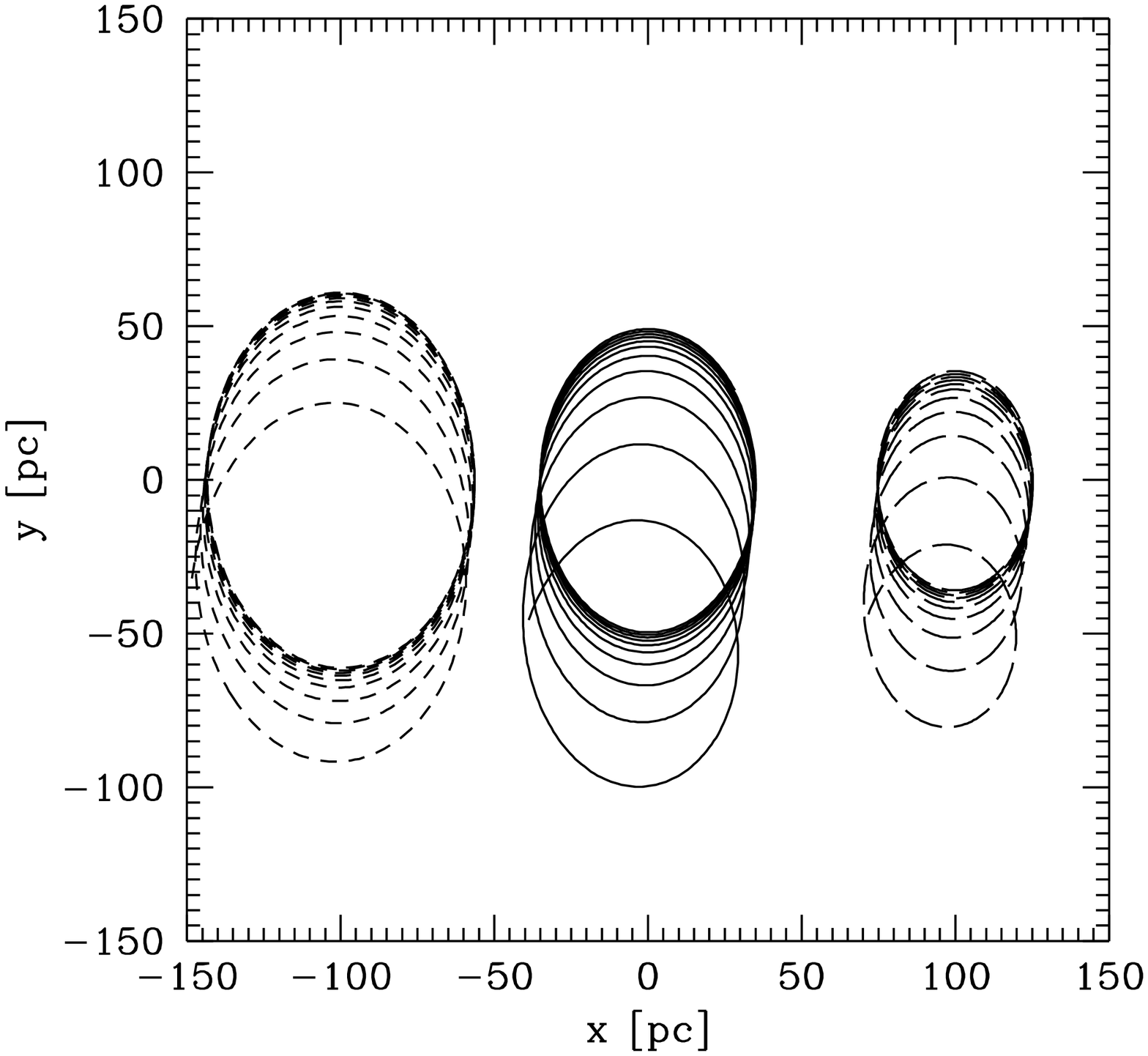} 
    \epsfxsize=8cm
    \epsfysize=8cm 
    \epsffile{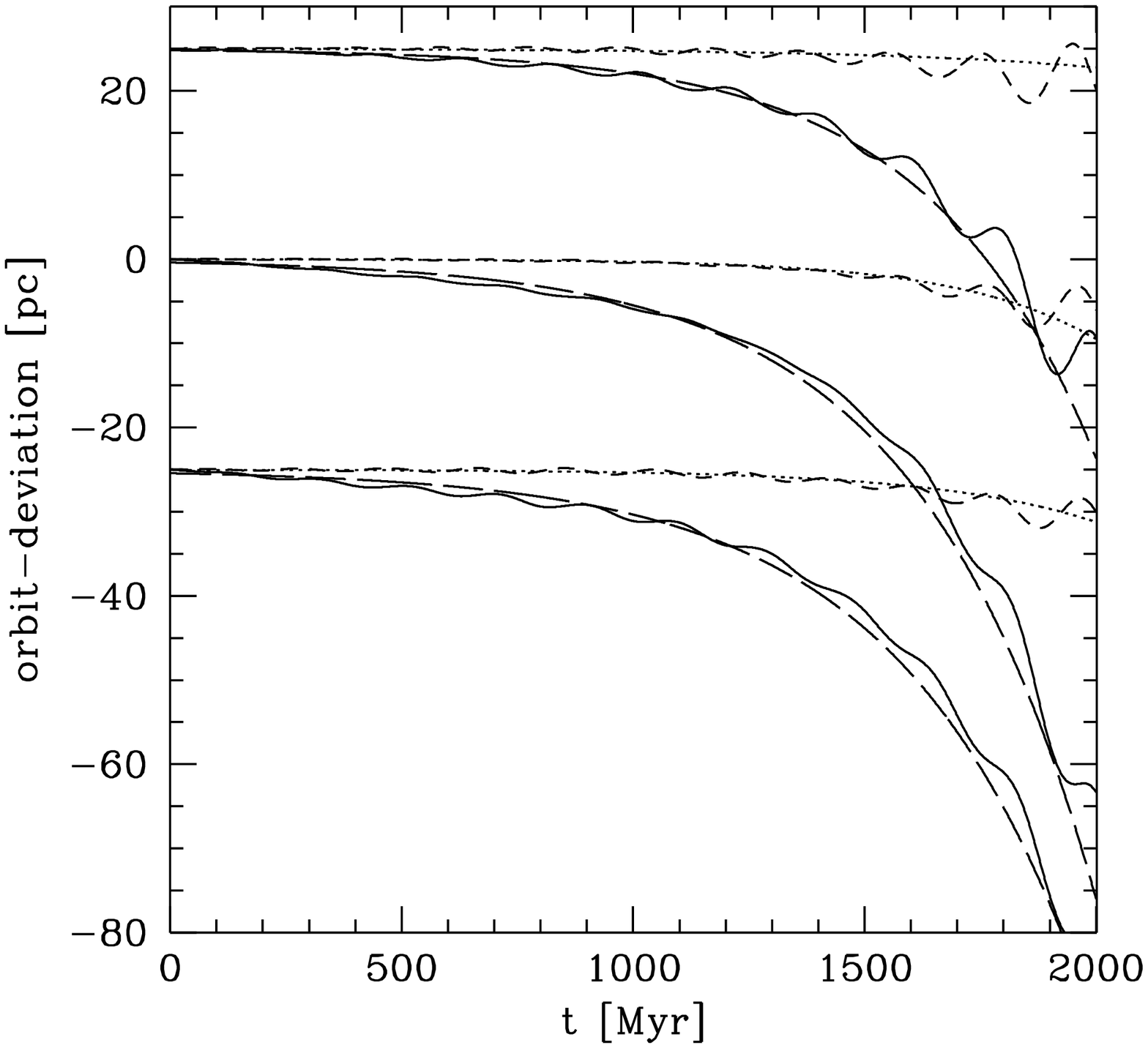}
    \caption{Upper: Orbits (in the co-moving cluster frame, for
      $t\leq2$~Gyr) of three particles in the simulation with
      $a=50$~pc and $\rho/\rho_{0}=0.1$.  For clarity two of the
      orbits have been shifted in the $x$-direction by $\pm100$~pc.
      One clearly sees the growing deviation from simple epicyclic
      motion.  Lower: $x$ (short dashed) and $y$ (solid) residuals
      (i.e. coordinates with the epicyclic solution subtracted) for
      the above three particles as functions of time.  The dotted
      ($x$) and long dashed ($y$) lines show fitting functions for an
      exponential growing solution.  Again for clarity the curves of
      two particles are shifted by $\pm25$~pc.}
    \label{fig:fitxy} 
  \end{center}
\end{figure}

The effect of the exponentially growing term can be clearly seen in
Fig.~\ref{fig:fitxy}, which shows particle orbits from the simulation
with $a=50$~pc and $\rho/\rho_{0}=0.1$ up to about $t_{diss}$.
Evidently, the exponential instability is much stronger in the
$y$-coordinate than in the $x$- coordinate; this is to be expected, by
eq.(\ref{eq:xiimg}).  The result is that the expansion of the system
takes place mainly in the direction along the orbit around the
galactic centre. 

\begin{figure}
  \begin{center} 
    \epsfxsize=8cm 
    \epsfysize=8cm
    \epsffile{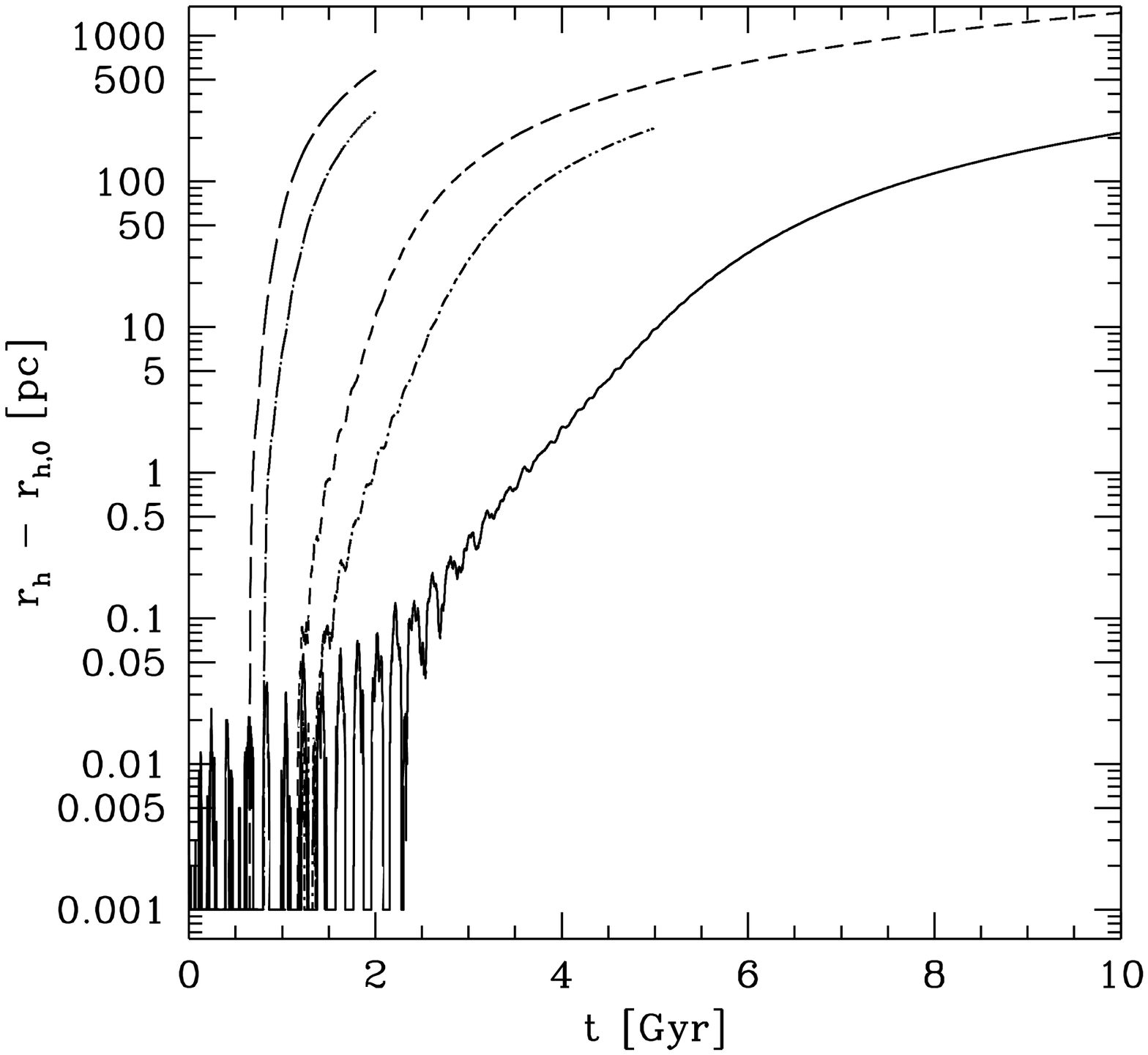} 
    \epsfxsize=8cm
    \epsfysize=8cm 
    \epsffile{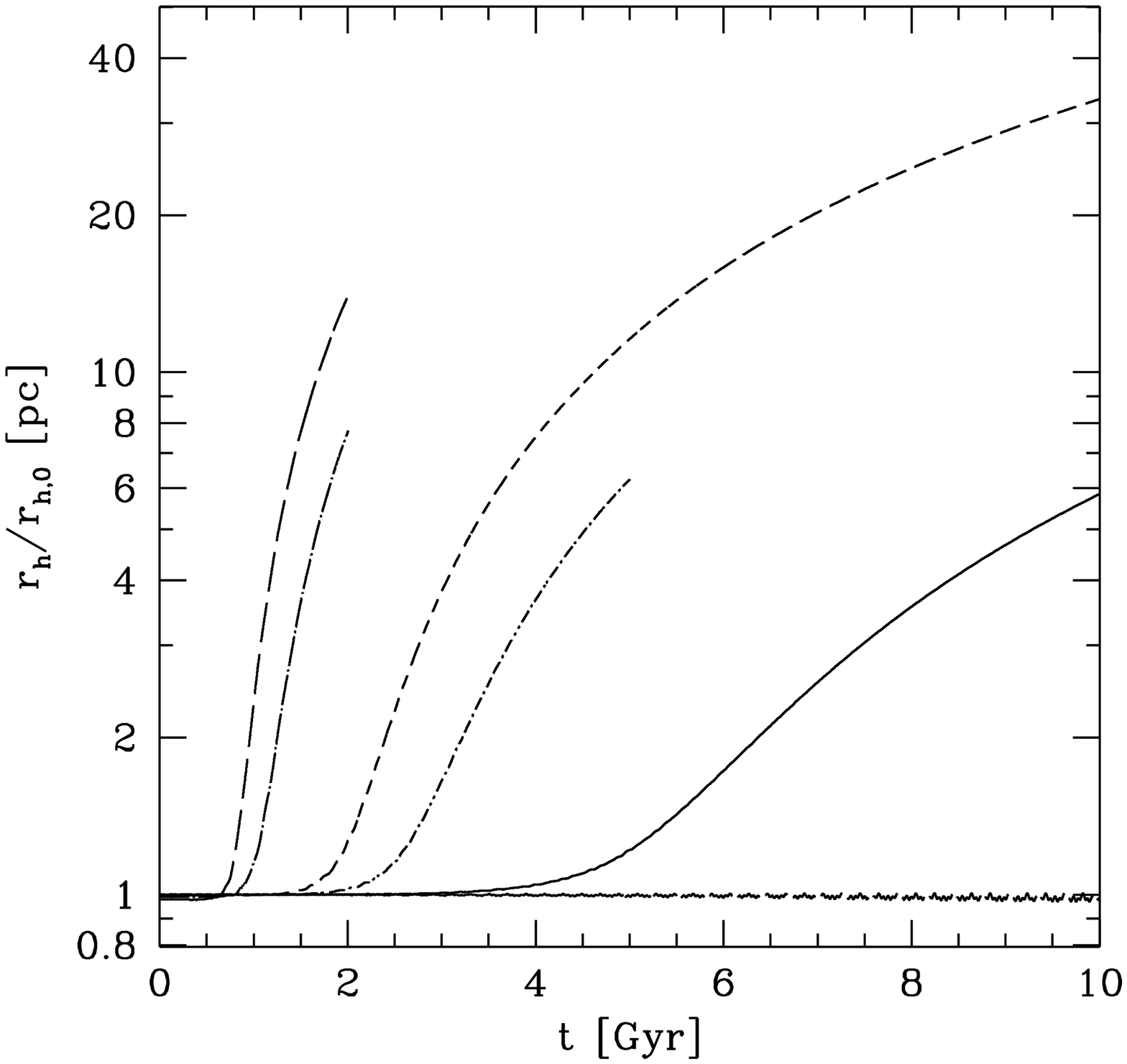}
    \caption{Upper: Difference between the current half-mass radius
      $r_{h}$ and the half-mass radius at the beginning of the
      calculations $r_{h,0}$.  All values are taken from simulations
      with $a=50$~pc and varying density (from left to right: 1.0,
      0.5, 0.1, 0.05 and 0.01 times the background density).  Lower:
      Ratio of $r_{h}$ and $r_{h,0}$ for the same calculations.  (The
      almost horizontal plot is a test calculation in which
      self-gravity was neglected in both the initial conditions and in
      the run.  For this model the only evolutionary mechanism is the
      inexactness of the epicyclic approximation.) The phase of
      exponential growth is clearly visible.  If the half-mass radius
      gets very large then the self-gravity of the system is
      negligible, and one sees the effects of epicyclic drift: the
      curves change to a linear growth.}  
    \label{fig:expgrowth} 
  \end{center}
\end{figure}

While Fig.~\ref{fig:fitxy} shows the effect of the exponential
instability on individual orbits, Fig.~\ref{fig:expgrowth} shows its
influence in the expansion of the system as a whole.  At very early
times it is hidden in the random noise of the simulation, or indeed by
the epicyclic motion itself.  Its presence is most clear at
intermediate times; later on, when the self-gravity of the system
becomes negligible, the expansion is sub-exponential.

\begin{figure}
  \begin{center} 
    \epsfxsize=8cm 
    \epsfysize=8cm
    \epsffile{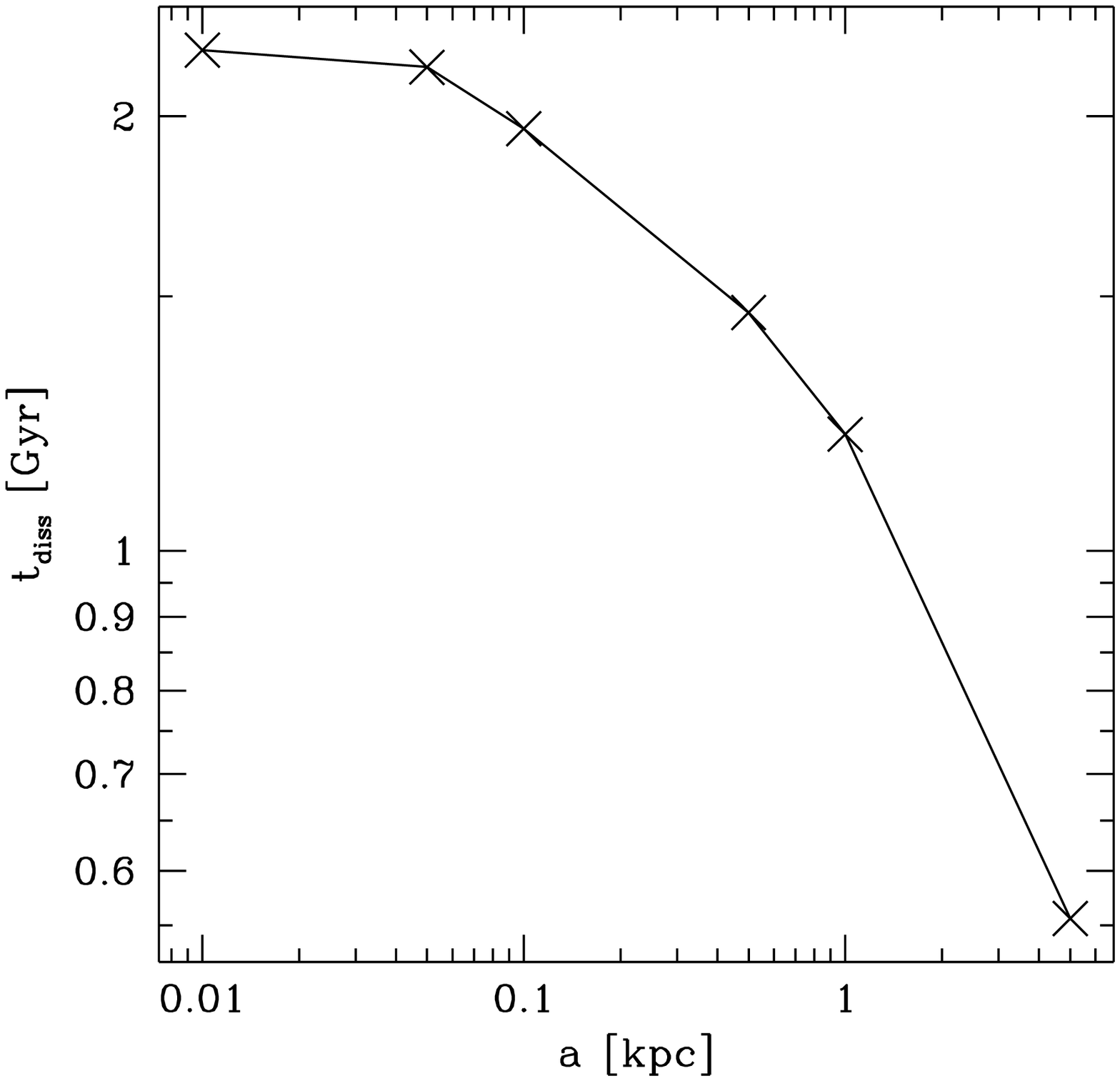} 
    \caption{Dissolution time as a function of the size $a$ of the
      system.  The density is kept constant at $\rho/\rho_{0} = 0.1$.} 
    \label{fig:diss2} 
  \end{center}
\end{figure}

We also investigated whether there is a relation between the size of
the system and the time of dissolution, but as one can see in
Fig.~\ref{fig:diss2} there is no obvious analytical relation.  The
fact that small systems (small $a$) have longer dissolution times than
large systems can be understood by noting that the epicyclic
approximation becomes less accurate for larger systems.

\subsection{Appearance of the systems}

\begin{figure}
  \begin{center}
    \epsfxsize=8cm 
    \epsfysize=8cm 
    \epsffile{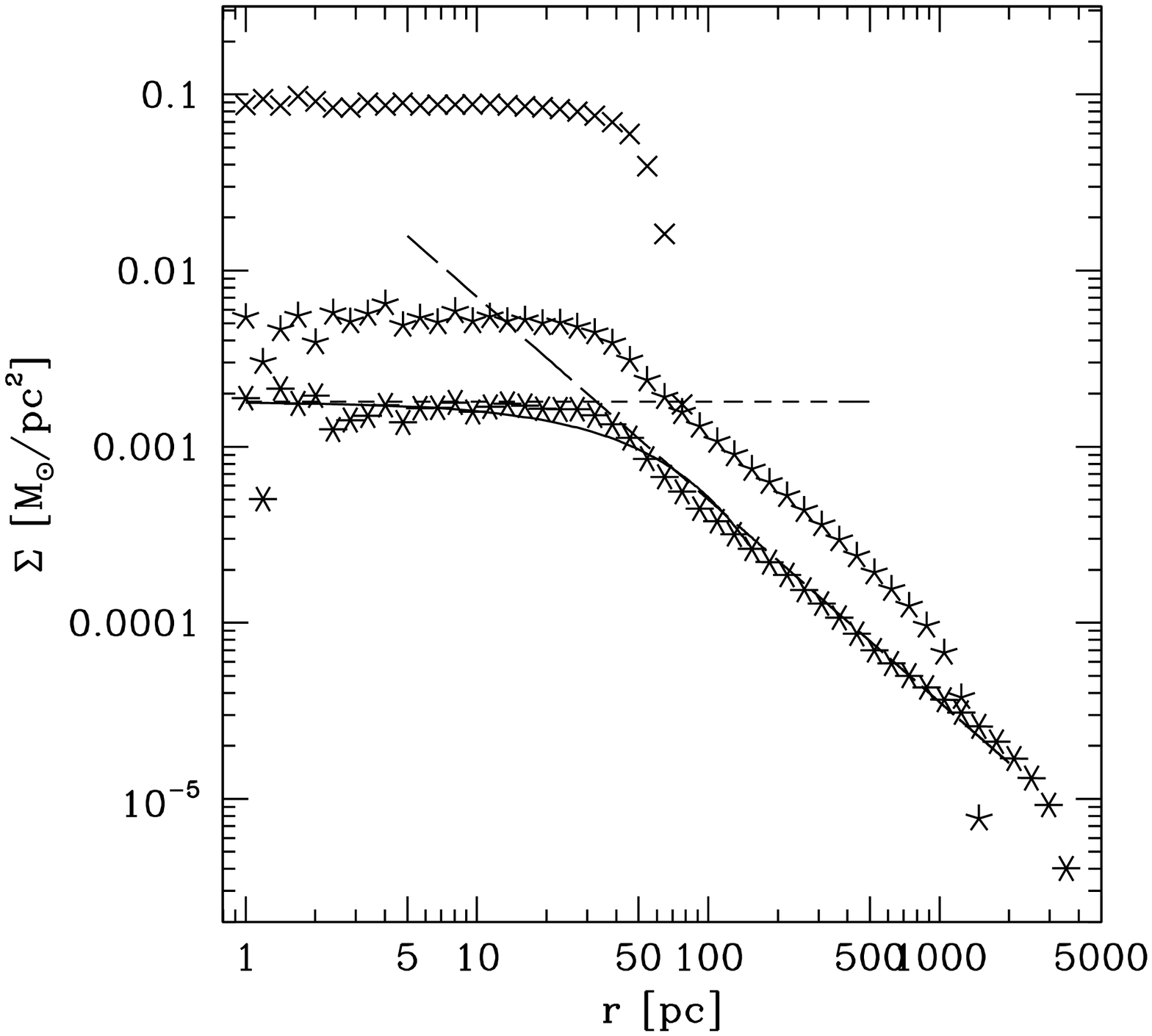}
    \caption{Surface density profile of the simulation with $a=50$~pc
      and $\rho/\rho_{0}=0.1$ at times $t=1$~Gyr (crosses), $t=5$~Gyr
      (five-pointed stars) and $t=10$~Gyr (six-pointed stars) measured
      in concentric radial bins centred on the object and looking
      along the $Z$-axis.  Also shown are the fitting lines for the
      $t=10$~Gyr data.  Short dashed line is the horizontal fit for
      the inner part, long dashed line the power law fit for the outer
      part and the solid curve is an exponential fit.}
    \label{fig:u03}
  \end{center}
\end{figure}

Though the initial conditions are highly idealised, it is interesting
to look briefly at how the appearance of such a system evolves as it
disperses.  Even after the system starts to disperse, forming tidal
tails along its orbit, we still find a measurable density enhancement
at the position of the unbound cluster.  In Fig.~\ref{fig:u03} 
we show the measured surface density in one of the runs at $t=1$, $5$
and $10$~Gyr.  The surface density distribution resembles that of an
object with a large core radius.  At late times the surface density is  
almost flat until nearly $100$~pc and then it turns over to a shallow
power law with an index of approximately $-1.6$.  Fitting an
exponential gives an exponential scale-length of approximately
$100$~pc.

The kinematics of the model depend on the frame in which it is viewed,
and are complicated by its initial rotation.  At $t=0$ it can be seen,
by setting $\lambda=0$ in eqs.(\ref{eq:epimot1})--(\ref{eq:epimot3}),
differentiating, and setting $t=0$, that $\dot x = -\kappa^{\prime}
y / \xi$, $\dot y = \xi \kappa^{\prime} x$ and $\dot z = - z_{0}
\nu^{\prime} \sin \phi$.  Thus along the $x$- and $y$-directions the
system appears to rotate, though with different angular velocities.
In the $z$-direction the velocity dispersion, $\sigma_{z}$, follows the
dispersion of $z_{0}$, i.e.\ the thickness in the $z$-direction.  We
also see that, at a given distance $r_{x}$ from the $x$-axis, the
dispersion in $\dot x$ is given by $\sigma_{x}^{2} = \langle \dot
x^{2} \rangle = \kappa^{\prime 2} \langle y^{2} \rangle / \xi^{2} =
\kappa^{\prime 2} r_{x}^{2} / (2 \xi^{2})$, with a similar result for
$\sigma_{y}^{2}$.  As the time of disruption (as defined here) is
approached, the profiles of $\sigma_x$ and $\sigma_y$ approach that 
for $\sigma_{z}$, which itself changes little from that at $t=0$.  The
evolution of $\sigma_{x}$ is faster, however, because the orbits diffuse
transverse to this line of sight more quickly (Fig.~\ref{fig:fitxy}).

In the galactocentric frame at $t = 0$ one sees from
eqs.(\ref{eq:coordtrafo1})-(\ref{eq:coordtrafo6}) 
that the corresponding results are $\dot X = Y ( \Omega -
\kappa^{\prime} / \xi)$ and $\dot Y = - \Omega D + (X-D) (- \Omega +
\xi \kappa^{\prime})$, while $\dot Z = \dot z$.  Again there appears to
be rotation (with different angular velocities) in the $X$ and $Y$
directions, but for $\dot X$ it can be shown from
eqs.(\ref{eq:epifreqold1}) and (\ref{eq:eqiparanew3}) 
that the angular velocity vanishes if $\rho = 0$.  As with the
velocity dispersions in the rotating frame, we find that all three
velocity dispersions become comparable as the time to disruption
approaches.  Beyond the original radius of the system, however, the
profiles of $\sigma_{x}$ and $\sigma_{y}$ increase with projected
radius, because of galactic rotation. 

The system which is shown in Fig.~\ref{fig:u03} has a total initial
mass of about $660$~M$_{\odot}$.  At $t=10$~Gyr the central
line-of-sight velocity dispersion is about $1$~kms$^{-1}$ and the core
radius is approximately $100$~pc.  If one followed the arguments of
Mateo (1998) to calculate the virial mass (assuming the object is
bound), using his formula 
\begin{eqnarray}
  \label{eq:virmass}
  M_{\rm vir} & \approx & 167 \beta r_{\rm core} \sigma^{2}_{\rm
  los}, 
\end{eqnarray}
with $\beta=8$, the result is roughly $10^{5}$~M$_{\odot}$.

\section{Conclusions and discussion}
\label{sec:disc}

We have studied the evolution of a special class of unbound systems
orbiting at a constant distance from the centre of an isothermal
galaxy model.  Initially the systems have uniform density within an
ellipsoidal region.  They move according to epicyclic theory (modified
for the self-gravity of the system), with guiding centre at the centre
of the ellipsoid.  These systems are exact self-consistent equilibrium
models within a steady tidal field.

While these equilibrium models are believed to be new in the context
of a tidally limited stellar system, they have properties in common
with Freeman's models of a barred galaxy \citep{fre66,bin87}.  There
too the effective potential inside the system is quadratic in a
uniformly rotating frame.  Freeman's models are the two-dimensional
limit of a flattened ellipsoid.  It would be interesting to explore
this link further. 

Our simulations show that, depending on their initial density and
radius, the systems we have constructed survive with only modest
expansion for a surprisingly long time.  When our results are scaled
to a circular orbit of radius $10$~kpc with a circular velocity of
$220$~kms$^{-1}$, a system of semi-minor axis $50$~pc and density
$0.2$ times that of the background galaxy survives for up to several
Gyr (i.e.\ about $20$ Galactic orbits).  For small systems the
dependence on initial density, and the direction in which the
dispersal takes place, can be understood on the basis of the modified
epicyclic theory.  For large systems the lifetime may be limited by
the accuracy of epicyclic theory.  Even though the systems gradually
disperse, there are measurable density enhancements even after a
Hubble-time. 

Our results have interesting parallels with those of \citet{kle03},
even though the initial conditions and galactic field are completely
different.  They investigated the survival of an unbound clump of
particles (with an initial Gaussian velocity distribution) in the
inner part of the dark halo of a dwarf satellite.  As long as the dark
matter halo is not cusped, they too found that the system disperses
only on long time-scales (longer than a Hubble-time).  The longevity
of their models relies partly on the relatively cold initial
conditions and partly on the near-harmonic potential of the core of
the host galaxy.  Our models are also cold initially (in the sense
that the velocity dispersion in $x-$ and $y-$ at a given point is
zero), and the potential is approximately harmonic inside the model
(eqs.(\ref{eq:epidiff1})--(\ref{eq:epidiff3})). 

Idealised models, such as those discussed in this paper, have the
advantage that they are suitable for analytical study, which can
provide greater understanding than the compilation of results from
simulations.  An example of this is our result on the dependence of
dissolution time on density.  The drawback of idealised models is that
the results cannot be applied directly to real systems, and can only
be suggestive.  On the other hand a study of Fig.\ref{fig:lagrad}
suggests that the role of the special initial conditions may be
confined to the phase in which the Lagrangian radii do not evolve by a
large factor.  The subsequent evolution (e.g.\ as shown in late-time
plots in Fig.\ref{fig:u03}) may be applicable more widely.

The systems in nature which most closely correspond to ours are open
clusters.  Shortly after the tidal radius of an evolving cluster has
shrunk to zero, we expect that the remnant consists of an
inhomogeneous spatial distribution of stars with a preponderance of
retrograde rotation, in nearly circular motion in the disc of the
galaxy.  Our models are initially homogeneous, with pure retrograde
rotation.  They nevertheless suggest that it would be interesting to
perform similar simulations with more realistic initial conditions.
If these also exhibited the long dissolution times we find for some
models, this would suggest that it may be possible to detect the
remnants of dissolved star clusters even now.

Other subsystems within galaxies (satellite galaxies and globular
clusters) differ further.  Their galactic orbits are not nearly
circular, as a rule, and the resulting time-dependent tide presumably
causes faster dissolution, which is enhanced also by disk shocking.
For this reason, we consider that the dissolution time we find (when
suitable scaled to the orbital period) is an upper limit to the
lifetime of such systems.

\begin{acknowledgements}
MF acknowledges financial support through DFG-grant FE564/1-1 and
KR1635/5-1.  We want to thank R. Spurzem, C. Theis and P. Kroupa for
fruitful discussions of the results.  We thank R.H. Miller for
suggesting the connection with Freeman's models.
\end{acknowledgements}

\end{document}